\documentclass[twocolumn,tighten,times]{aastex62}

\usepackage[T1]{fontenc}

\newcommand\ergs{erg~s$^{-1}$}

\begin{document}

\title{A Wind-Disk Self-Irradiation Model For Supercritical Accretion}
\shorttitle{A Wind-Disk Self-Irradiation Model For Supercritical Accretion}

\author[0000-0001-8018-5348]{Yuhan Yao}
\affiliation{Cahill Center for Astrophysics, California Institute of Technology, MC 249-17, 1200 E California Boulevard, Pasadena, CA, 91125, USA}

\author[0000-0001-7584-6236]{Hua Feng}
\affiliation{Department of Astronomy, Tsinghua University, Beijing 100084, China}
\affiliation{Department of Engineering Physics, Tsinghua University, Beijing 100084, China}

\correspondingauthor{Hua Feng}
\email{hfeng@tsinghua.edu.cn}

\begin{abstract}

Optical emission from actively accreting X-ray binaries is dominated by X-ray reprocessing on the outer disk. In the regime of supercritical accretion, strong radiation will power a massive wind that is optically thick and nearly spherical, and will occult the central hard X-rays from irradiating the outer disk. Instead, thermal emission from the wind will act as a new source of irradiation. Here, we construct a self-irradiation model, in which the inner disk (within the wind photosphere) is completely blocked by the wind, the middle part (between the wind photosphere and scattersphere) is heated by the wind directly, and the outer disk (beyond the wind scattersphere) is heated by photons leaving the scattersphere.  The model can adequately fit the UV/optical SED of NGC 247 X-1, a candidate source with supercritical accretion, while the standard irradiation model fails to produce a self-consistent result. The best-fit parameters suggest that the source contains a stellar mass black hole with an accretion rate roughly 100 times the critical value. Remarkably, the UV/optical fitting predicts a wind photosphere that is well consistent with X-ray measurements, although it is an extrapolation over 3 orders of magnitude in wavelength. This implies that supercritical accretion does power a massive wind and the UV/optical data are useful in constraining the wind structure. 
 
\end{abstract}


\section{Introduction}
\label{sec:intro}

During the outburst of low-mass X-ray binaries, the optical emission is found to be dominated by disk self-irradiation \citep{vanParadijs1994,Gierlinski2009}, i.e.,  thermal emission from an X-ray heated outer disk.  The optical emission on an irradiated disk is a function of the X-ray luminosity and the disk structure and size \citep{Frank2002}. Thus, optical study of actively accreting X-ray binaries could help constrain the geometry of the accretion flow and evolution history of the binary system.  

Disk irradiation is also found to dominate the optical emission in ultraluminous X-ray sources \citep[ULXs;][]{Kaaret2009,Tao2011,Roberts2011,Grise2012,Soria2012,Soria2012a,Tao2012,Sutton2014}, which are 2-3 orders of magnitude more luminous than Galactic X-ray binaries \citep[for a review, see][]{Kaaret2017}.  The majority of ULXs are argued to be powered by supercritical accretion onto stellar compact objects \citep[e.g.,][]{Gladstone2009,Middleton2015}.  Discovery of the ultraluminous pulsars provides smoking gun evidence for this scenario \citep{Bachetti2014, Fuerst2016, Israel2017,Israel2017a,Carpano2018,Sathyaprakash2019}. However, the physics for supercritical accretion is still poorly understood. Both analytical analysis \citep{Shakura1973, Lipunova1999,King2003,Poutanen2007,Shen2016} and numerical simulations \citep{Ohsuga2011,Jiang2014,Sadowski2016} predict a unique signature that, driven by strong radiation,  an optically thick and nearly spherical wind will be launched and encircle an optically thin\footnote{optically thin to absorption but thick to scattering.} funnel at the center.  The presence of the wind shapes the emergent X-ray emission: hard X-rays from the central disk may be geometrically beamed when they propagate through the central funnel \citep{Middleton2015,Weng2018}, and the radiation from the disk beneath the optically thick wind is reprocessed to be thermal emission on the wind photosphere \citep{Soria2016,Urquhart2016,Feng2016,Zhou2019,Tao2019}.  Such a geometry suggests that the traditional physical picture of disk irradiation, where a point-like X-ray source at the center illuminates the outer disk, is no longer valid due to occultation by the wind.  

However, irradiation models based on the standard disk geometry were often used to fit the spectral energy distributions (SEDs) of ULXs \citep[e.g.,][]{Gierlinski2009,Grise2012,Tao2012} or compare with the photometric colors \citep[e.g.,][]{Copperwheat2007,Patruno2008,Madhusudhan2008}. These models assume an irradiating source at the center of a standard disk and part of its power is intercepted by the outer disk. This is valid only if the hard X-ray emission is fairly isotropic. However, for systems with extremely high accretion rates, the hard X-rays from the central funnel could be considerably beamed so that the outer disk sees very little amount of them. In these extreme cases, the irradiating source becomes the thermal emission from the optically thick wind instead of the central hard X-rays. 

Some luminous and very soft sources are good candidates of these systems. \citet{Zhou2019} assembled a sample of such sources and found that they could be best explained as due to supercritical accretion onto stellar mass black holes or neutron stars, featuring a very soft ($kT_{\rm bb} = 0.05-0.4$~keV), Eddington-limited thermal emission as a signature of the optically thick wind. The inferred accretion rate is around 100--500 times the critical value based on a radiation hydrodynamic (RHD) model \citep{Meier1982_PaperII,Meier1982_PaperIII}. Thus, for these systems, the standard irradiation geometry will fail due to occultation by the wind. Irradiation models taking into account the presence of wind and/or a different disk structure under supercritical accretion have been proposed \citep{Vinokurov2013,Ambrosi2018}. These models assume a geometry in which the disk and wind are stacked at characteristic radii.  Here, we propose a new irradiation model based on the RHD wind model of \citet{Meier1982_PaperII} so that the radiative transfer in the wind can be considered. This is non-trivial because the thermal photons from the wind photosphere will be multiply scattered until they leave the scattersphere, which is much larger in size than the photosphere.  This will cause a non-negligible effect on the irradiation.  We will describe the model in Section \ref{sec:model}, compare the model with data in Section \ref{sec:fit}, and discuss the results in \ref{sec:discuss}. 

\section{Model}
\label{sec:model}

Detailed description of the base model can be found in \citet{Meier1982_PaperII}, Chapter 13 of \citet{Meier2012}, and the Appendix of \citet{Zhou2019}.  Here, we elaborate how irradiation works in this frame. We define the dimensionless compact object mass $m = M/M_\sun$, Eddington luminosity $L_{\rm Edd} = 1.5 \times 10^{38}m$~\ergs, critical accretion rate $\dot{M}_{\rm Edd} = L_{\rm Edd} / 0.1c^2$, and the dimensionless accretion rate $\dot{m} = \dot{M} / \dot{M}_{\rm Edd}$.  

The wind is assumed to launch above a slim disk close to the radius where the advective energy loss is comparable to the radiative energy loss.  This is also the radius where the total luminosity approaches the Eddington limit. Radiation pressure will expel some or most of the accretion matter into the wind, with the rest (at least $\dot{m} \approx 1$) continuing to accrete and power the wind. In the case of $\dot{m} \gg 1$, we assume that the majority of the accretion matter goes into the wind with the same density, drift velocity, pressure, and temperature at the advective radius. These set the inner boundary conditions of the wind.  Then, the wind develops from the inner boundary to infinite by solving the RHD equations \citep{Meier1982_PaperII}, with a transition from acceleration to free expansion.  

In this work, as we are focusing on the disk irradiation, the two most important radii are the photosphere radius $r_\ast$ and the scattersphere radius $r_{\rm sc}$.  The temperature of thermal emission is determined at the photosphere. Between $r_\ast$ and $r_{\rm sc}$, the wind is optically thin to absorption but optically thick to scattering, and its radiative flux will heat the underlying disk directly. Beyond $r_{\rm sc}$, the irradiating source can be regarded as the surface of the scattersphere. Prescriptions of $r_\ast$, $r_{\rm sc}$, and other structural and thermal properties of the wind, can be found in the Appendix of \citet{Zhou2019}.

\begin{figure}
\includegraphics[width=\columnwidth]{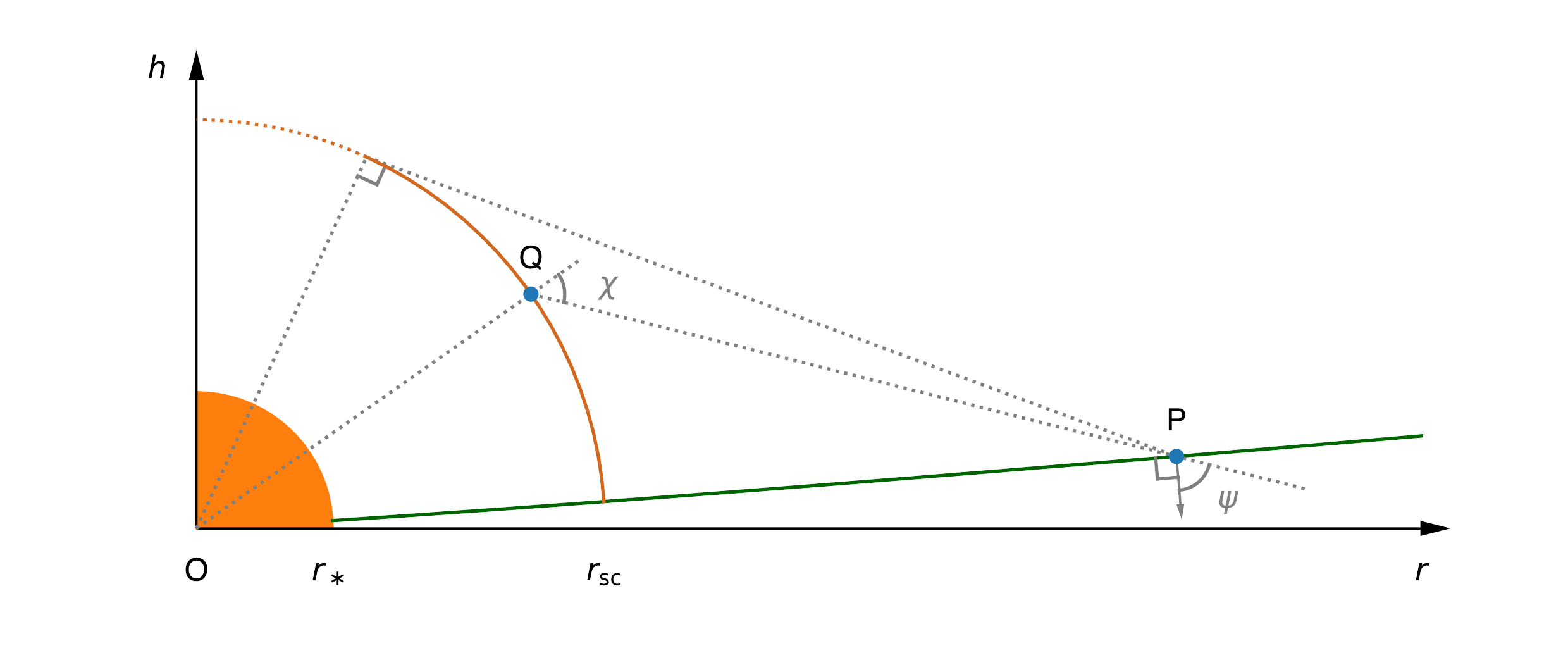}
\caption{A schematic drawing of the self-irradiation model. The photosphere radius of the wind is denoted as $r_\ast$, within which the disk cannot be seen by an observer. The scattersphere of the wind is denoted as $r_{\rm sc}$, beyond which the photons no longer interact with the wind. The green curve indicate a standard disk. Between $r_\ast$ and $r_{\rm sc}$, the disk is heated by the local radiation in the wind directly. Above $r_{\rm sc}$, the disk is heated by photons on the portion of scattersphere that is visible to the disk element.
\label{fig:irr_model}}
\end{figure}

A schematic drawing of irradiation model is shown in Figure~\ref{fig:irr_model}.  A standard disk structure \citep{Shakura1973} is adopted beyond $r_\ast$ as the radius is large enough. Due to the high opacity of the disk, irradiation only changes the surface temperature of the disk but will not alter the disk structure, which is determined by its mid-plane temperature \citep{Dubus1999}. The half height of a standard disk at $r > r_\ast$ is
\begin{equation}
h =  (8.6 \times 10^{3} \; {\rm cm}) \, m^{9/10} \, \dot{m}^{3/20} \, \alpha^{-1/10} \, (r / r_{\rm isco})^{9/8},
\end{equation}
where $r$ is the radius and $r_{\rm isco} \equiv 6GM/c^2$ is the radius of the innermost stable circular orbit.  

Between $r_\ast$ and $r_{\rm sc}$, the disk is heated by the radiative flux in the local wind directly, with an irradiating temperature 
\begin{equation}
T_{\rm irr}(r) = \left[ \frac{f_{\rm wind} (r) \; (1 - \beta)}{\sigma} \right]^{1/4}  \, ,
\label{eq:inside}
\end{equation}
where $\beta$ is the albedo, $f_{\rm wind} (r)$ is the radiative flux in the wind at radius $r$, and $\sigma$ is the Stefan-Boltzmann constant. Above the photosphere, the effective absorption optical depth goes below unity while the scattering optical depth is still high.  In this region, the wind radiative flux at radius $r$ can be calculated assuming radiative diffusion from the photosphere following
\begin{equation}
f_{\rm wind} (r) = \frac{4 \tau_{\rm es}(r) r_\ast^2 \sigma T_\ast^4}{3 \tau_{{\rm es}, \ast} r^2} \, ,
\end{equation}
where $\tau_{\rm es}(r) \approx \kappa_{\rm es} \rho(r) r$ is the scattering optical depth of the wind at radius $r$. At the scattersphere, where $\tau_{\rm es} (r_{\rm sc}) = 1$, the wind luminosity freezes at
\begin{equation}
L_{\rm bb} = \frac{16 \pi r_\ast^2 \sigma T_\ast^4}{3 \tau_{{\rm es},\ast}} =  \frac{3}{4}L_{\rm Edd} \, .
\end{equation}

Above $r_{\rm sc}$, as shown in Figure~\ref{fig:irr_model}, the total power that a disk annulus can receive is an integral of the flux from the part of the scattersphere that is visible to the annulus. The flux at the scattersphere is $f_{\rm sc} = L_{\rm bb} / (4 \pi r_{\rm sc}^2)$. The flux per unit area intercepted by a disk element at radius $r$ is
\begin{equation}
\sigma T_{\rm irr}^4 (r) = f_{\rm sc}
\int_{\theta_0}^{\theta_1} \int_{\phi_0}^{\phi_1} 
\frac{2 r_{\rm sc}^2 \sin\theta d\theta d\phi \cos\chi \cos\psi \, (1 - \beta) }{\pi \| { \rm QP} \|^2},
\label{eq:outside}
\end{equation} 
where $\theta$ and $\phi$ are the polar and azimuthal angle, respectively, $\chi$ and $\psi$ are indicated in Figure~\ref{fig:irr_model},  and $\| {\rm QP} \|$ is the distance from the scattersphere surface element to the disk annulus. Considering the visibility and thickness of the disk, the integral limits on the hemisphere are $\theta_0 = \arcsin [h(r_{\rm sc}) / r_{\rm sc}]$, $\theta_1 = \arccos(r_{\rm sc} / r)$, $\phi_0 = \arcsin[h(r_{\rm sc}) / (r_{\rm sc} \sin\theta)]$, and $\phi_1 = \pi - \phi_0$.

In addition to the irradiated flux, the viscous power for a standard disk is given by
\begin{equation}
\sigma T_{\rm vis}^4(r) = \frac{3G M \dot M}{8 \pi r^3 } \; .
\label{eq:vis}
\end{equation}
The effective temperature of the disk with irradiation can be obtained as
\begin{equation}
T_{\rm eff}^4(r) = T_{\rm vis}^4(r) + T_{\rm irr}^4(r) \; . 
\end{equation}

The disk within the scattersphere is covered by a Thomson thick wind and thus the photon flux will be reduced due to radiative diffusion in the wind. As a result, the observed flux from this part of disk is attenuated by a factor of $4 / [3 \tau_{\rm es}(r)]$.  Due to the same reason, photons from this part of the disk are redirected and become isotropic when they escape from the scattersphere.

\begin{figure}
\includegraphics[width=\columnwidth]{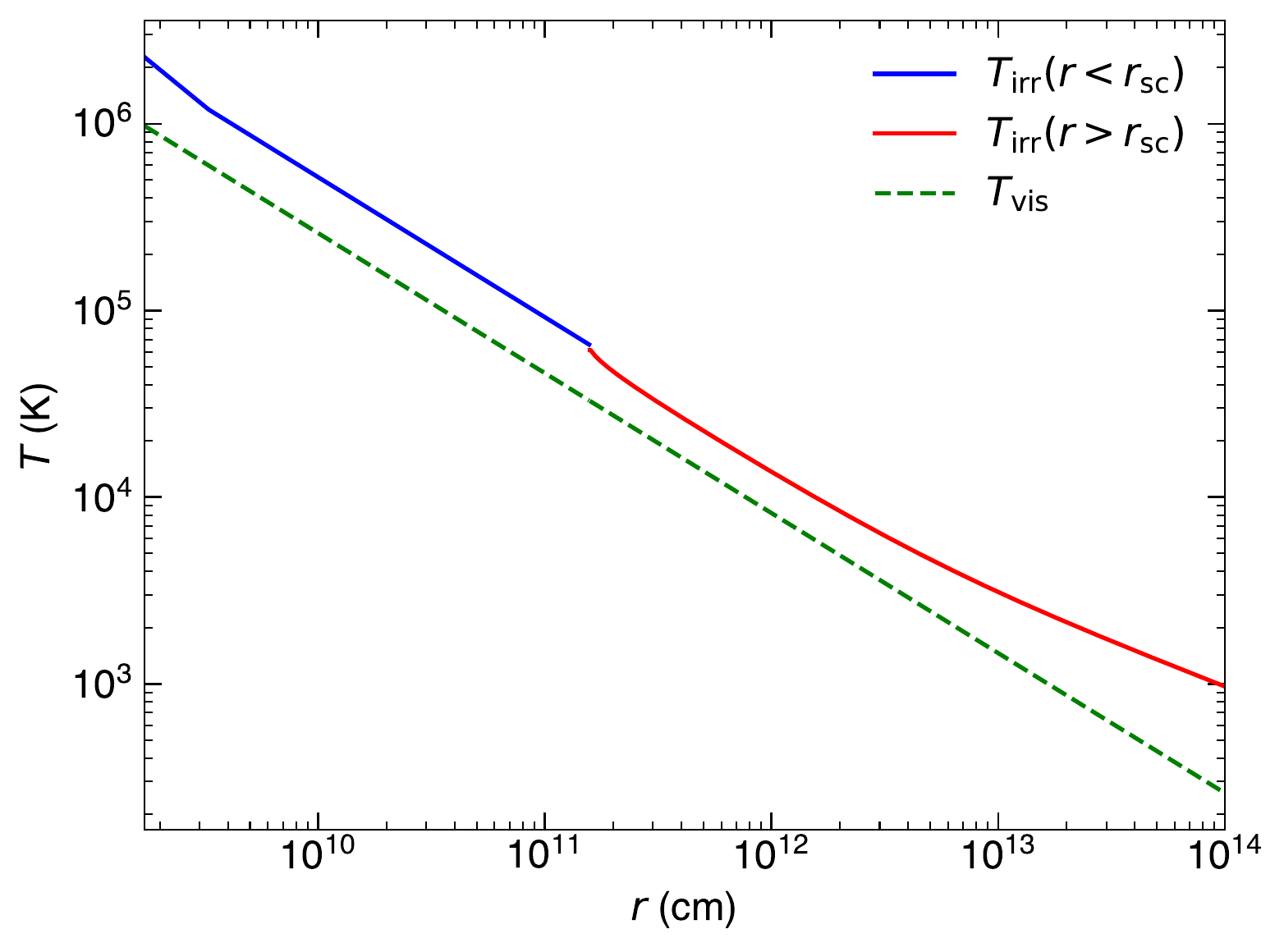} \\
\includegraphics[width=\columnwidth]{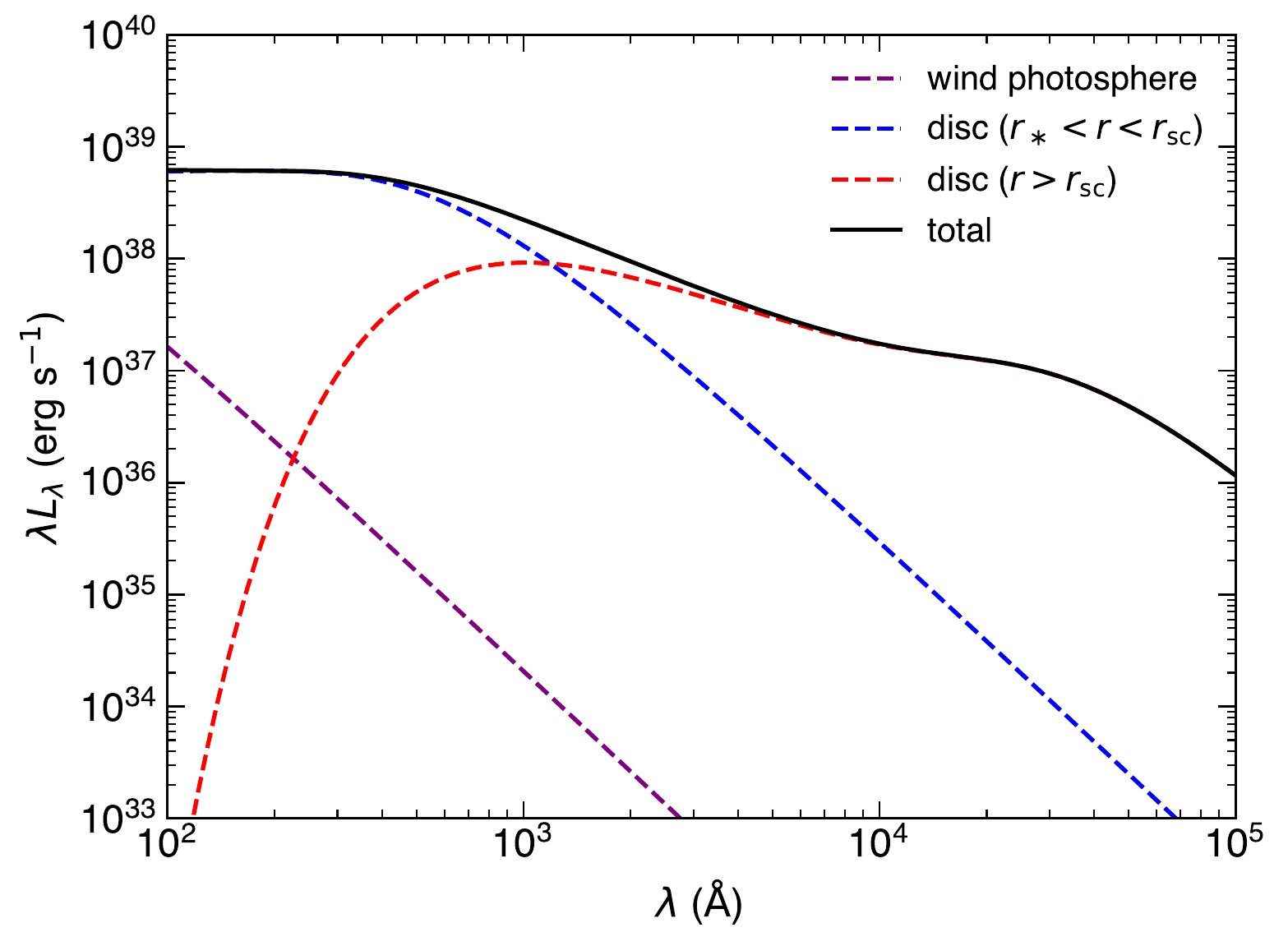}
\caption{Temperature profile and emergent spectrum in the UV/optical/IR band of the irradiation model given $m = 10$, $\dot{m} = 100$, $\alpha = 0.1$, $\beta = 0.7$, a disk outer radius $r_{\rm out} = 10^{14}$~cm, and a viewing angle ${\rm i} = 60^\circ$. The model predicts a photosphere radius $r_\ast = 1.7 \times 10^9$~cm and a scattersphere radius $r_{\rm sc} = 1.6 \times 10^{11}$~cm. We note that the viewing angle only affects the emission above $r_{\rm sc}$.
\label{fig:temp_spec}}
\end{figure}

In Figure~\ref{fig:temp_spec}, a temperature profile and an emergent spectrum of the outer disk are shown given the following parameters: $m = 10$, $\dot{m} = 100$, $\alpha = 0.1$, $\beta = 0.7$, a disk outer radius $r_{\rm out} = 10^{14}$~cm, and a viewing angle $i = 60^\circ$. As one can see, the disk temperature outside the wind photosphere is predominantly determined by irradiation. The reprocessed emission in the region $r_\ast < r < r_{\rm sc}$ dominates in the UV band, while emission beyond $r_{\rm sc}$ dominates in the optical and longer wavelengths. 

\begin{figure*}
\includegraphics[width=0.33\textwidth]{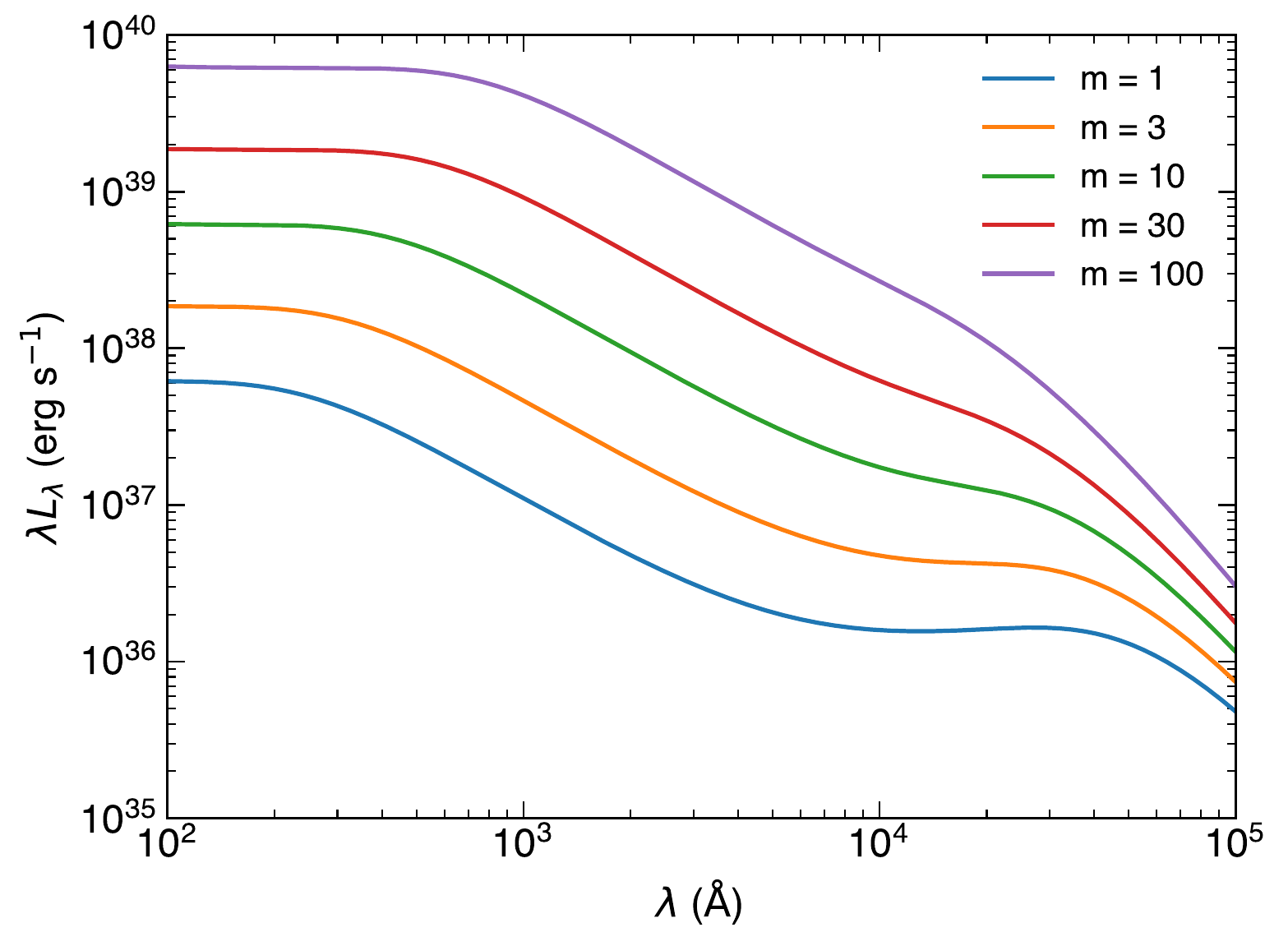}
\includegraphics[width=0.33\textwidth]{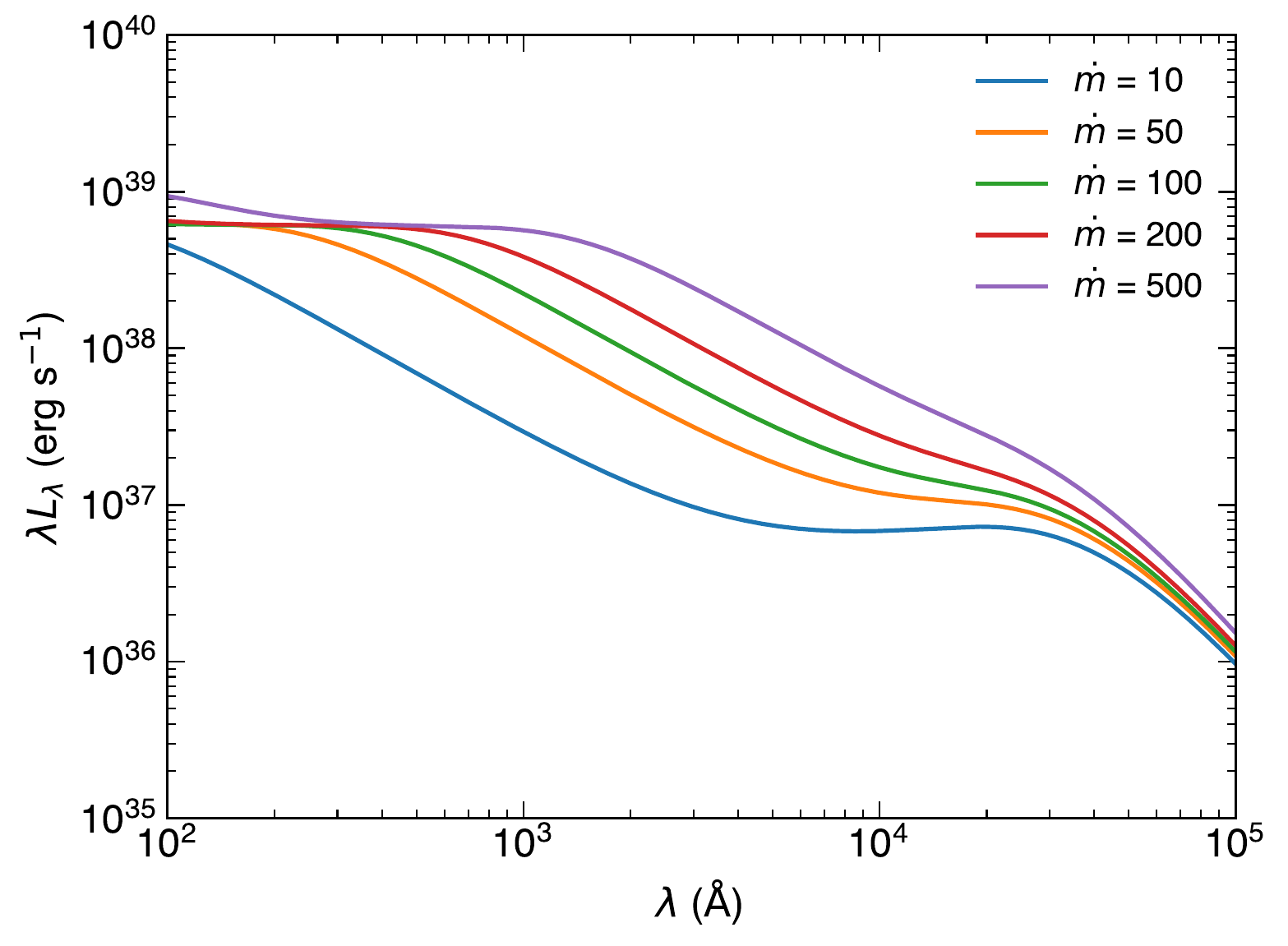}
\includegraphics[width=0.33\textwidth]{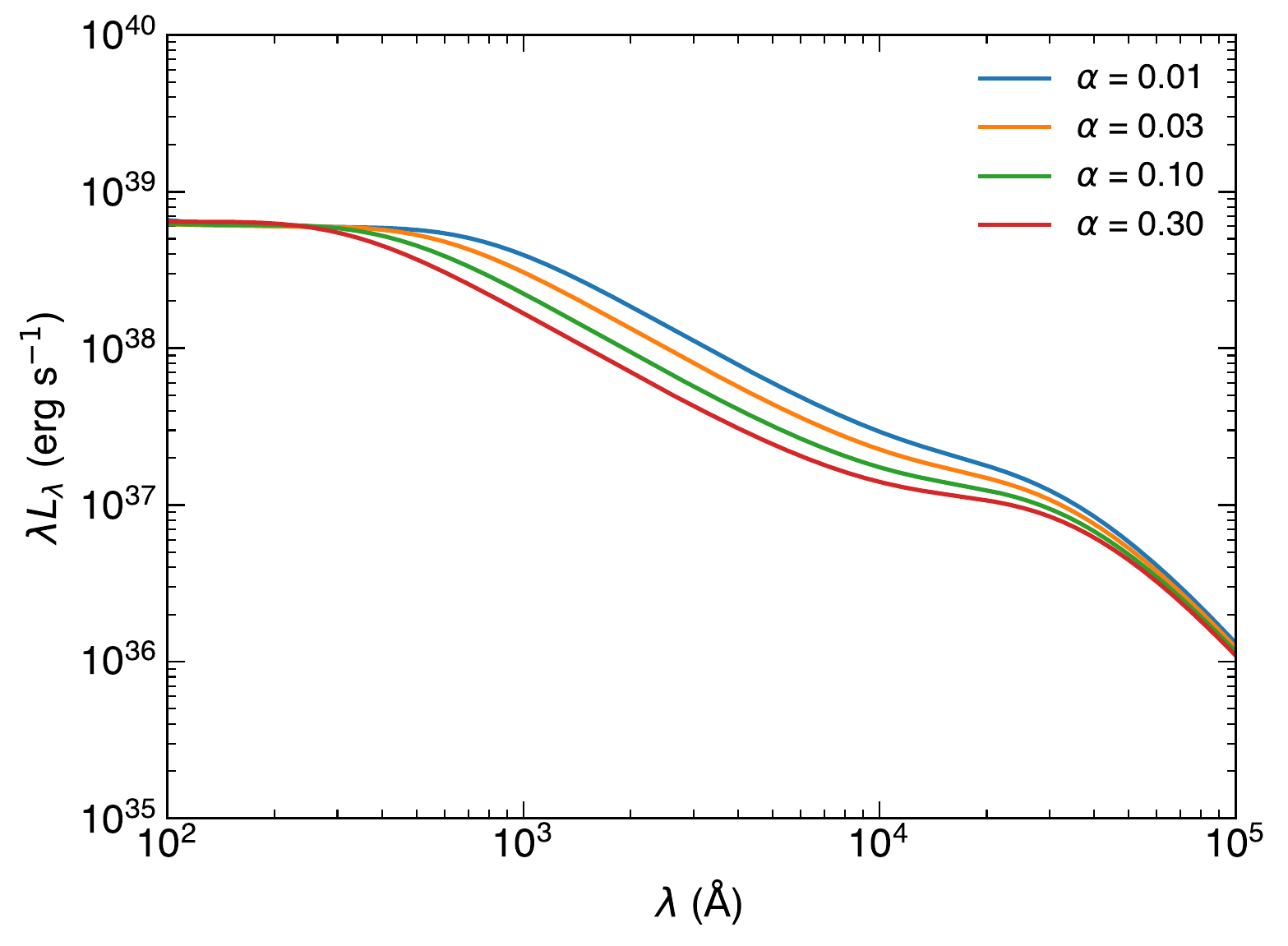}\\
\includegraphics[width=0.33\textwidth]{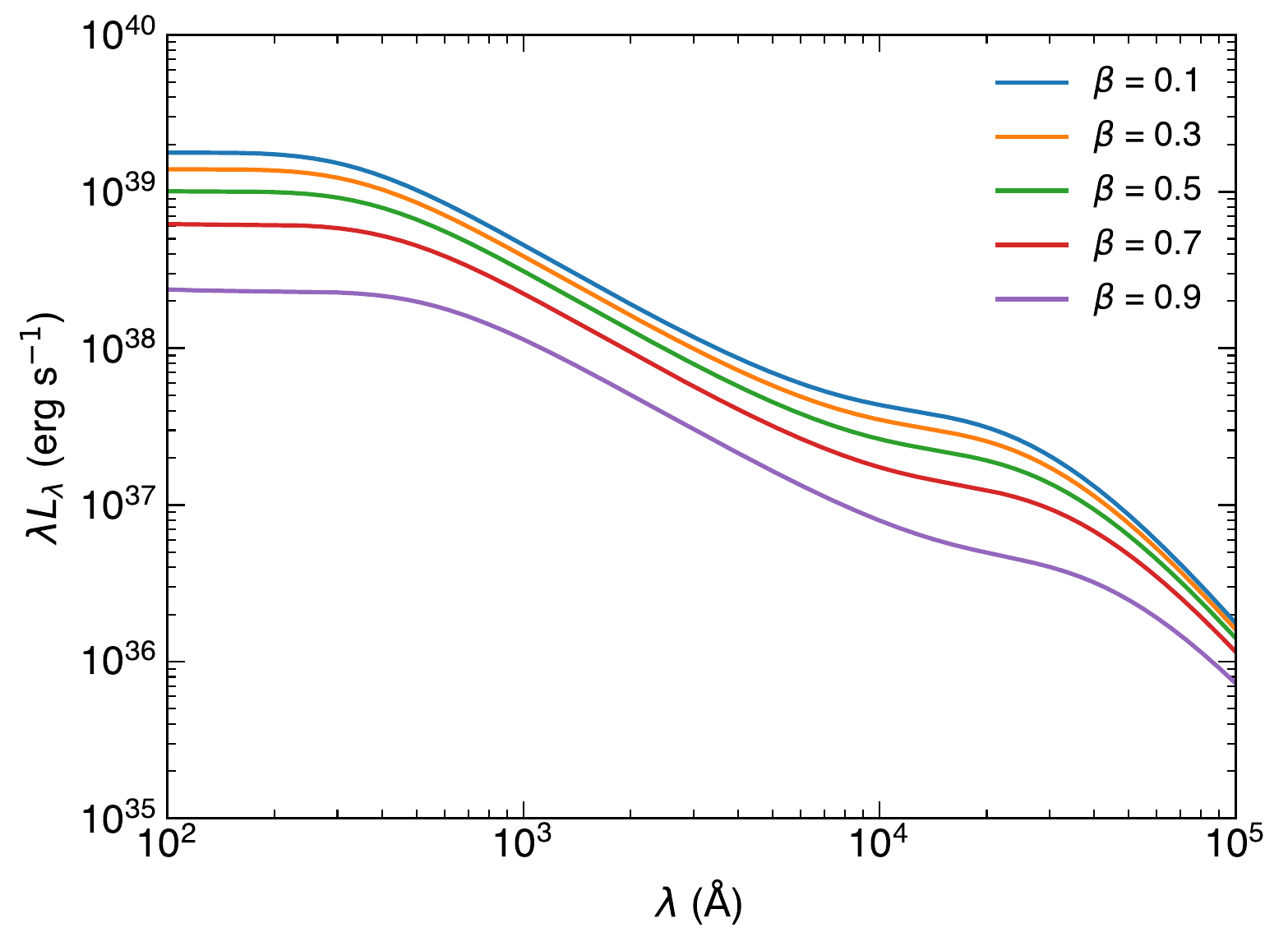}
\includegraphics[width=0.33\textwidth]{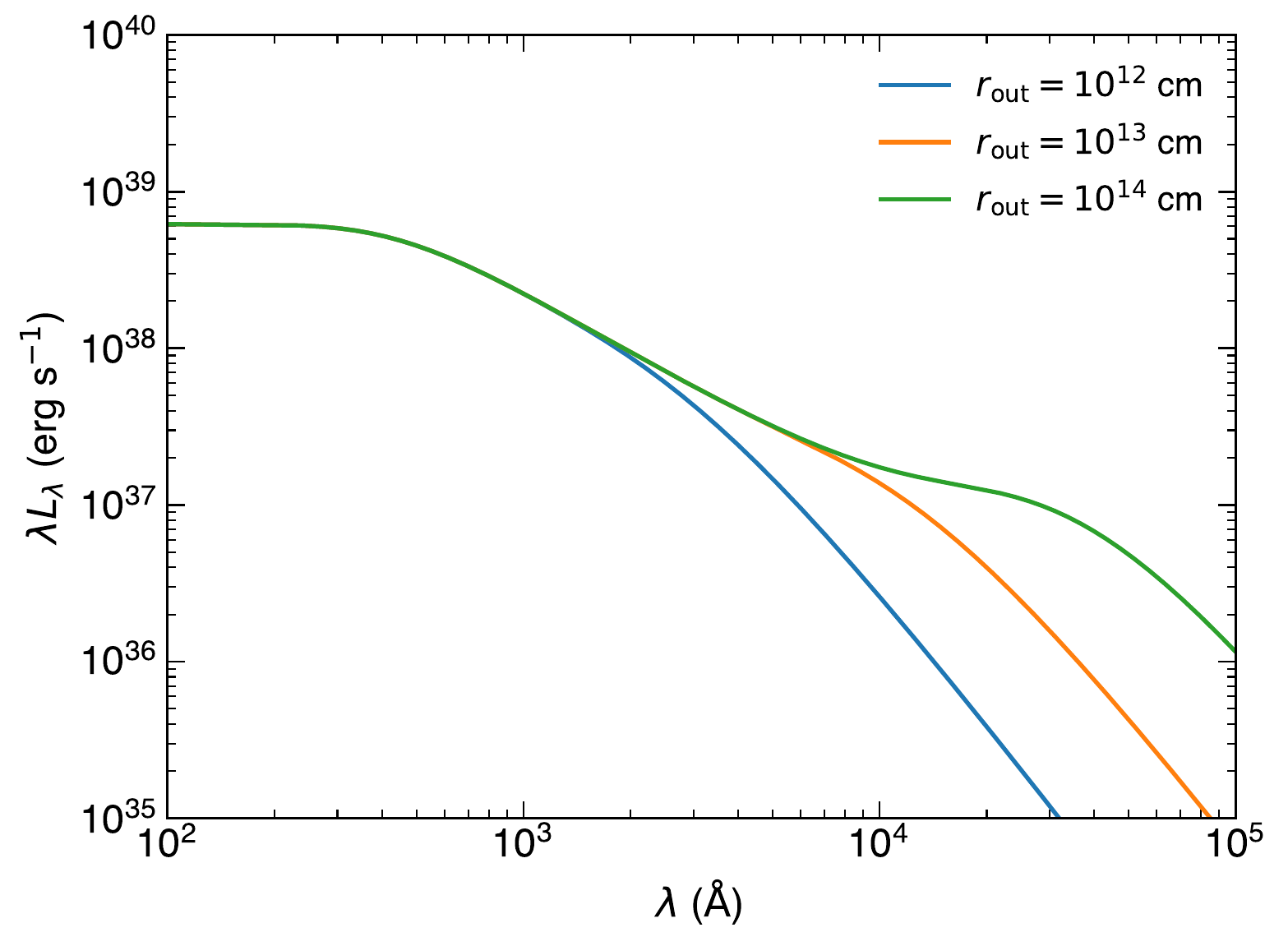}
\includegraphics[width=0.33\textwidth]{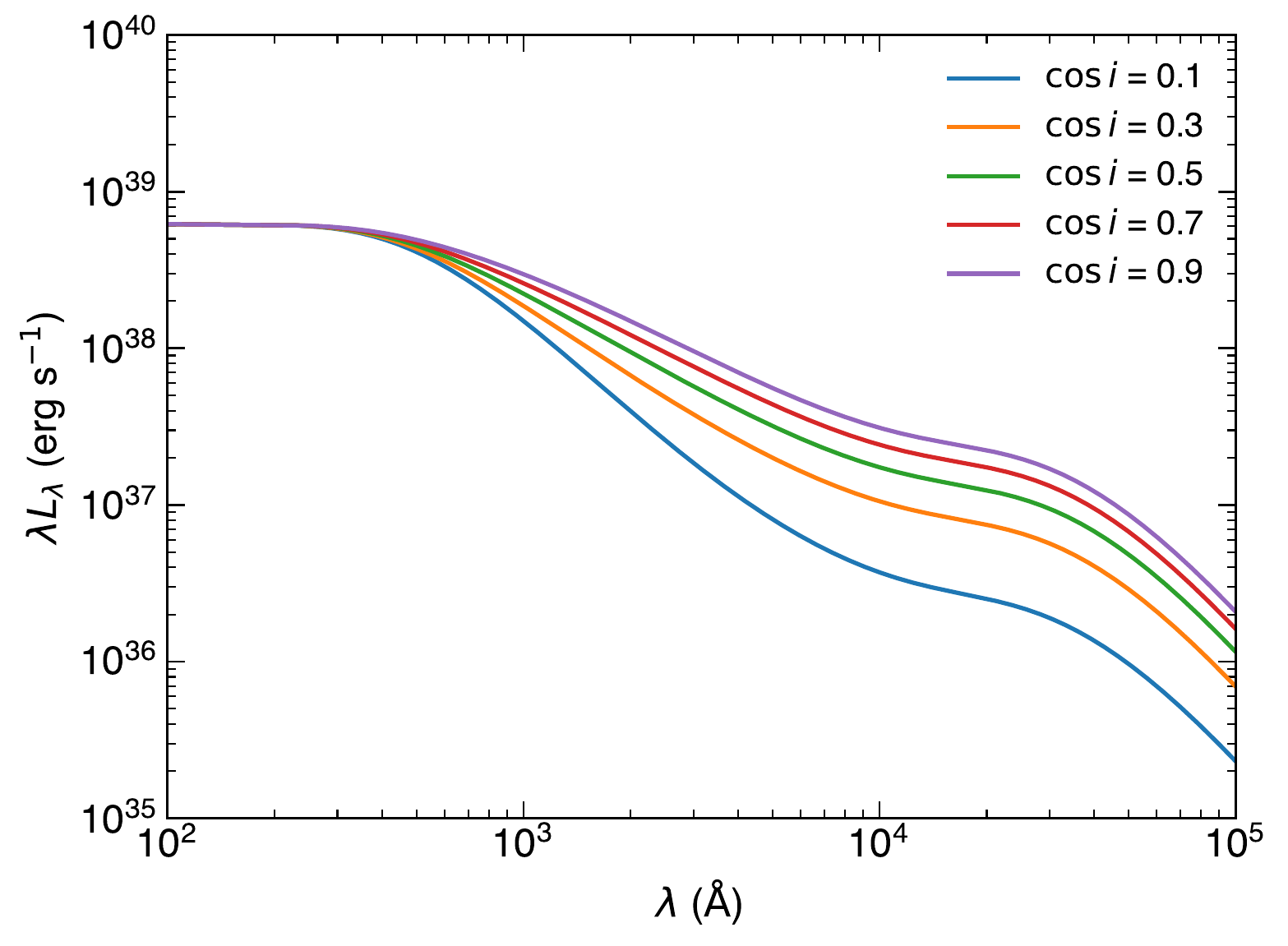}
\caption{Model spectra with different parameters. The base model assumes $m = 10$, $\dot{m} = 100$, $\alpha = 0.1$, $\beta = 0.7$, $r_{\rm out} = 10^{14}$~cm, and $i = 60^\circ$,  In each panel, models are generated with one of the parameters varying with values listed in the panel. 
\label{fig:para}}
\end{figure*}

To illustrate how the parameters affect the observed spectrum, we plot model spectra in Figure~\ref{fig:para} by varying one of the above parameters. The two parameters of major interest are the compact object mass and accretion rate, i.e., $m$ and $\dot{m}$. As one can see, $m$ mainly determines the total luminosity, but also controls the turnover wavelength in the UV band, where disk emission from inside and outside the scattersphere equals each other.  This turnover wavelength around $10^3$\AA\ is also a strong function of $\dot{m}$.  In addition, when $\dot{m}$ is extremely high, say $\gtrsim 500$, the Rayleigh-Jeans tail of the blackbody emission from the photosphere will become dominant in the UV band. Thus, the UV spectrum is the key to constraining the nature of the supercritical accretion via irradiation spectroscopy.  The infrared (IR) spectrum is also sensitive to some of the parameters. However,  due to the low flux density of the disk and relatively strong contamination from the companion star and stellar environment in IR, it is often complicated to interpret the IR spectrum.  

Among all parameters, $\alpha$ and $\beta$ have the smallest influence on the spectrum. A constant luminosity of the illuminating source (wind photosphere) can be understood as the reason why $\alpha$ is important for accretion but not here. Numerical simulations suggest that $\alpha = 0.01 - 0.1$ is reasonable for supercritical accretion but varies with radius \citep{Jiang2014}. $\beta$ mainly modifies the absolute flux scale and has little effect on the spectral shape.  $r_{\rm out}$ apparently determines the long wavelength cutoff of the spectrum. The inclination angle changes the observed flux ratio of the emission inside the scattersphere, which is isotropic, to that outside. Thus, in this work we fix $\alpha = 0.1$ and $\beta = 0.7$ because of the small degree of freedom in the data. To compare with the SED, one also needs to consider the extinction law and total amount of extinction. 

\section{SED fit}
\label{sec:fit}

The model assumes that there is no irradiation by hard X-rays from the central funnel. Thus, we want to choose objects that are very soft and luminous, with negligible hard X-ray emission. These sources are argued to be good candidates undergoing supercritical accretion \citep{Zhou2019}.  As mentioned above, UV spectrum is the key to testing the model. By searching the literature, there is only one source, NGC 247 X-1 (aka NGC 247 ULX), with such data available \citep{Tao2012,Feng2016}. The SED consists of a slitless UV spectrum and broadband photometry covering a wavelength range from 1350\AA\ to 15000\AA, measured with the Hubble Space Telescope (HST).  The data are adopted from \citet{Feng2016}, who found that the IR flux at wavelengths above 1 $\mu$m might have a higher photometric uncertainty or could be due to a secondary component. We thereby discard the two IR points and choose the UV and optical data from 1350\AA\ to 8000\AA\ in our SED fitting. A distance of 3.4~Mpc to NGC 247 is adopted \citep{Gieren2009}.

The fitting is done in two steps. First, a grid search is performed in the space of parameters of interest, including $\log m$ (from 0 to 2 with 20 points), $\log \dot{m}$ (from $\log 50$ to 3 with 20 points), $\log (r_{\rm out} / {\rm cm})$ (from 11 to 14 with 16 points), $\cos i$ (from 0 to 1 with 11 points), extragalactic $E(B-V)$ (from 0 to 0.3 with 30 points), and the extragalactic extinction law (one of the seven in the \texttt{PySynphot} package\footnote{Lim, P.~L., Diaz, R. ~I., \& Laidler, V.\ 2015, PySynphot User's Guide (Baltimore, MD: STScI), https://pysynphot.readthedocs.io/en/latest/. The extinction laws are named as \texttt{gal3},  \texttt{mwdense}, \texttt{mwrv21}, \texttt{mwrv4}, \texttt{lmc30dor}, \texttt{lmcavg}, and \texttt{smcbar} in the package.}). $\alpha$ and $\beta$ are fixed at their standard values as discussed above. A Galactic extinction with $E(B-V) = 0.018$ and $R(V) = 3.1$ (\texttt{gal3} in PySynphot) is always applied. Given the SED and measurement errors, $\chi^2$ is calculate at each node of the grid. 

Then, the parameters that produce the minimum $\chi^2$ are adopted as the initial parameters for the second step, where the Markov Chain Monte Carlo (MCMC) method is used to fit the SED utilizing a Python package \texttt{emcee}.\footnote{https://emcee.readthedocs.io} The extinction law \texttt{smcbar} is fixed in the MCMC fit, as in the grid search, it appears exclusively in the 155 sets of parameters that result in the smallest $\chi^2$.  This law is derived by observations of the Small Magellanic Cloud star-forming bar \citep{Gordon2003} with $R(V)=2.74$. It is not in conflict with the fact that NGC 247 may have a subsolar metallicity on average \citep{Davidge2006}, and that ULXs are more favored at subsolar metallicities \citep{Mapelli2010}. In addition to the measurement errors, an intrinsic uncertainty that is proportional to the flux (constant in magnitude) is assumed in the likelihood function, to account for possible systematics in the data and/or model. A uniform prior is assumed for each parameter. One hundred walkers are randomly generated around the initial parameters by adding a tiny Gaussian shift, and run for 1,000 steps. The posterior distribution after 400 steps is found to be stable and used to estimate the uncertainty of the parameters. 

\begin{figure}
\includegraphics[width=0.9\columnwidth]{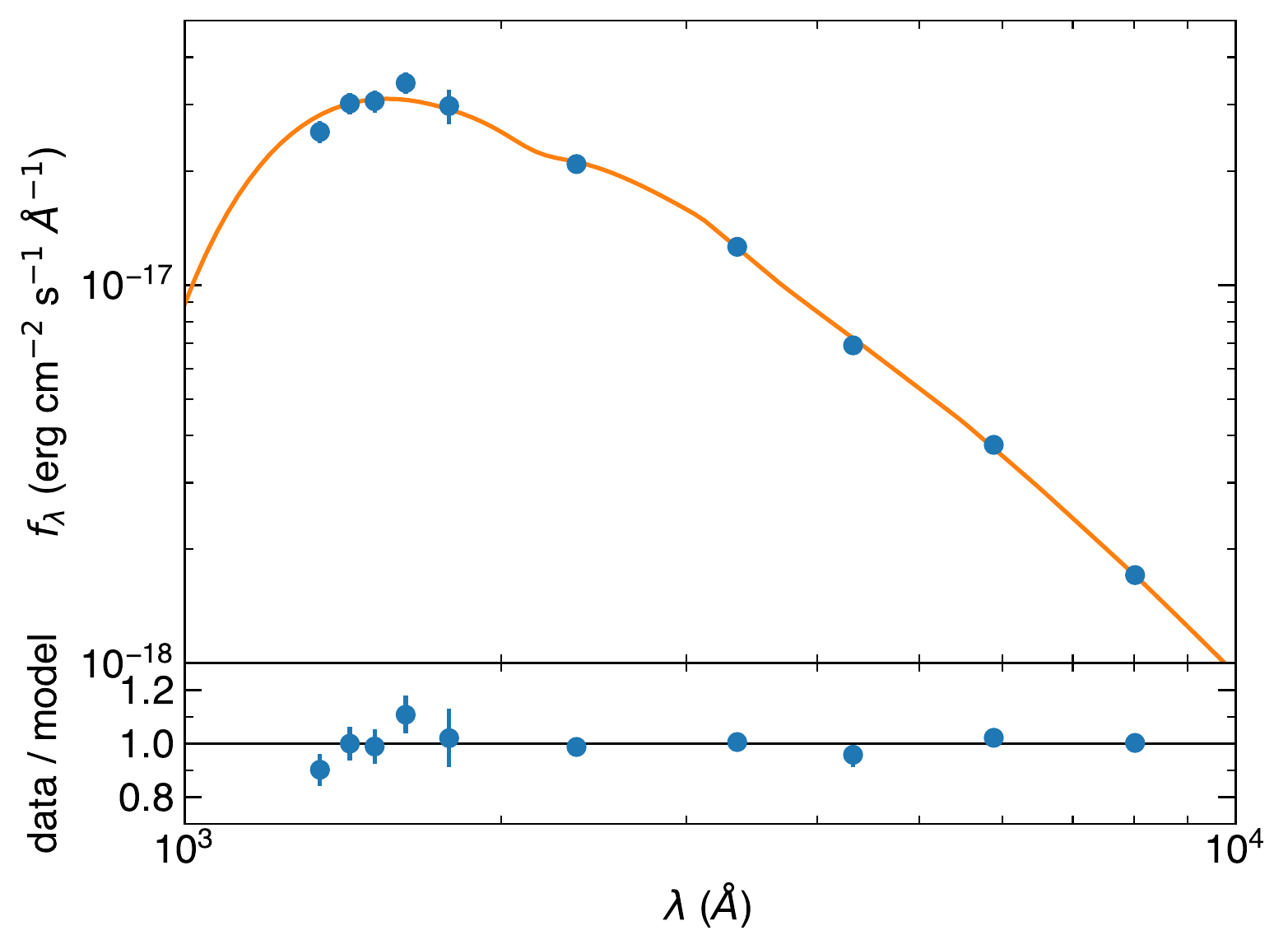}
\caption{Observed UV/optical SED of NGC 247 X-1 and the best-fit model.  
\label{fig:best-fit}}
\end{figure}

\begin{figure*}
\includegraphics[width=0.9\textwidth]{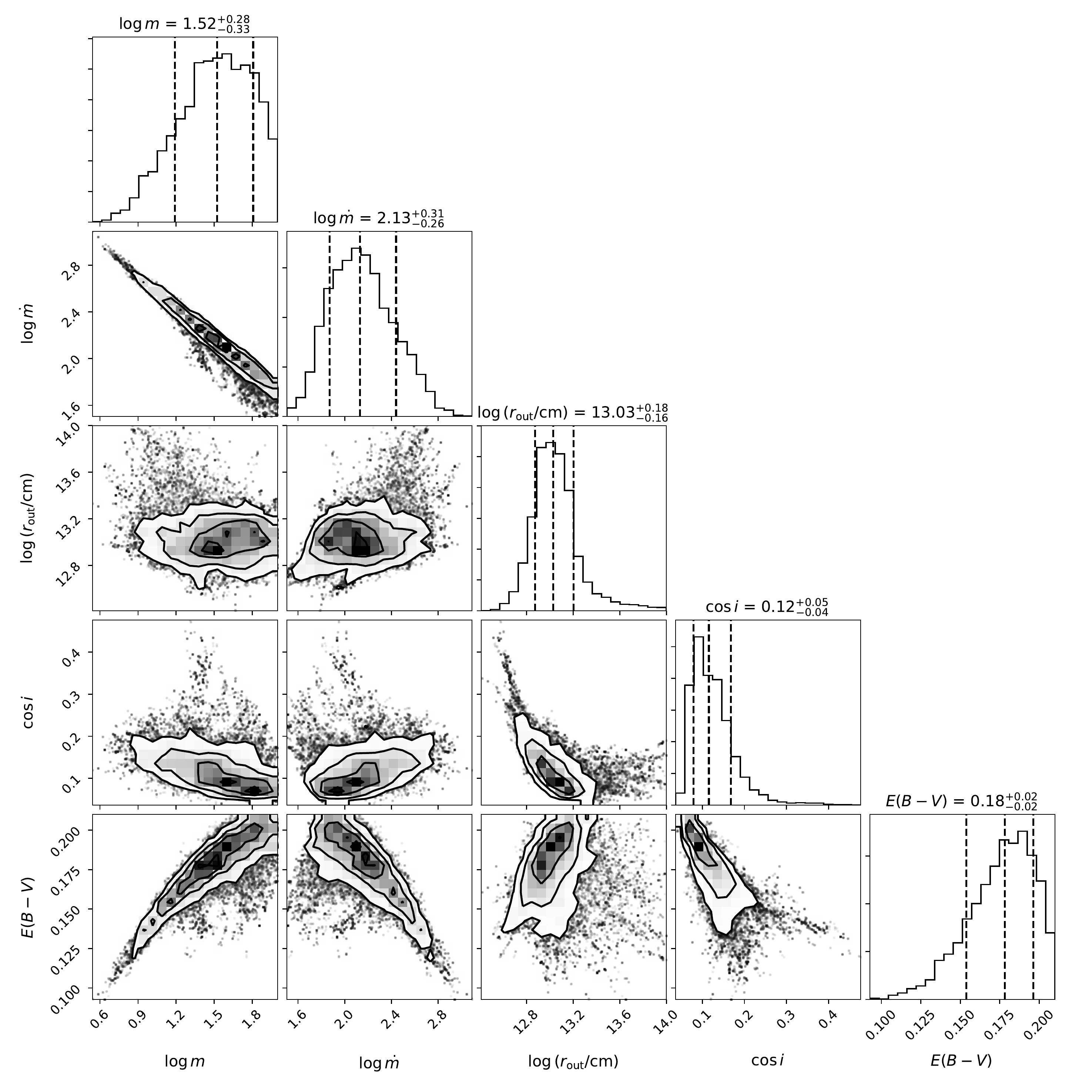}
\caption{Posterior distributions and bivariate correlations of the parameters after MCMC.
\label{fig:mcmc}}
\end{figure*}

The best-fit parameters with 1$\sigma$ uncertainties are: 
$\log m = 1.5_{-0.3}^{+0.3}$, 
$\log \dot{m} = 2.1_{-0.3}^{+0.3}$, 
$\log (r_{\rm out} / {\rm cm}) = 13.03_{-0.16}^{+0.18}$, 
$\cos i < 0.39$ ($3\sigma$), 
$E(B-V) = 0.178_{-0.025}^{+0.018}$, 
and the intrinsic scatter is around 0.16\%.  We note that the best-fit extinction in the host galaxy agrees with the mean extinction in NGC 247 ($E(B-V) = 0.18$) measured using Cepheids \citep{Gieren2009}.   The measurement errors have a min/median/max of 1.9\%/5.5\%/10.7\%, suggesting that the intrinsic scatter is tiny and negligible. The observed SED and best-fit model are plotted in Figure~\ref{fig:best-fit}. Their posterior distributions and bivariate correlations are shown in Figure~\ref{fig:mcmc}. As one can see, there is a degeneracy between the compact object mass and accretion rate. A more massive black hole with a lower accretion rate or a less massive black hole with a higher accretion rate can both fit the UV/optical SED. 

\begin{figure}
\includegraphics[width=0.9\columnwidth]{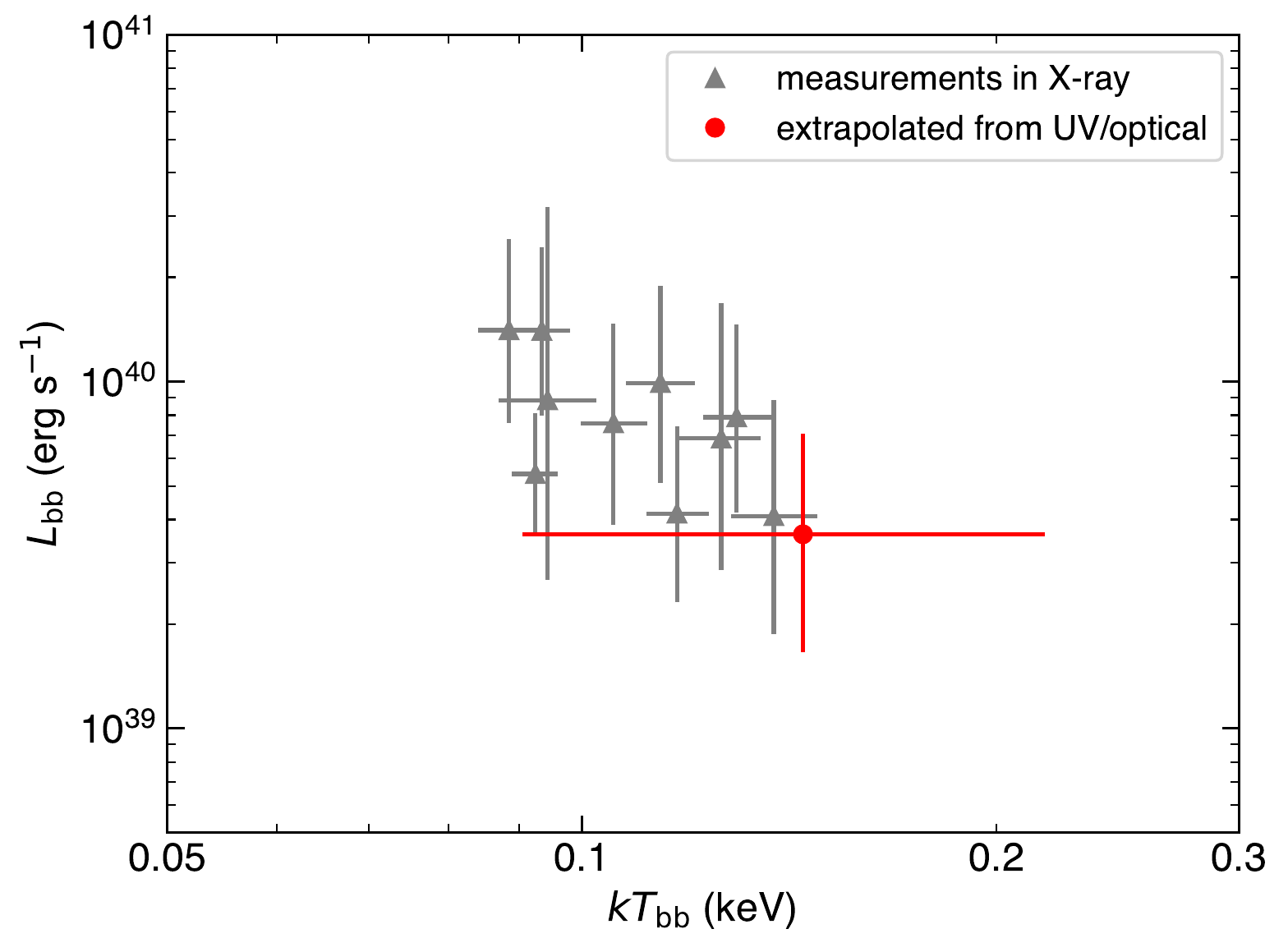}
\caption{Temperature and luminosity of the wind photosphere derived from UV/optical fitting (\textbf{red}) compared with measurements in the X-ray band (\textbf{gray}), which are adopted from \citet{Feng2016}.  They are well consistent with each other although there is an extrapolation over 3 orders of magnitude in wavelength. All errors are of 1$\sigma$. 
\label{fig:xray}}
\end{figure}

The X-ray emission of this source is dominated by a blackbody component, argued to arise from the photosphere of the wind driven by supercritical accretion. The photosphere itself is not directly seen in UV or optical (see Figure~\ref{fig:temp_spec} bottom), but its temperature and luminosity can be predicted by fitting the UV/optical SED.  This allows for an important test of the model.  We note that a joint fit is inappropriate, even if the observations are scheduled simultaneously, as dramatic variations in the X-ray band have been detected, with a timescale of as short as 200~s \citep{Feng2016}.  In Figure~\ref{fig:xray}, the luminosity and temperature of the blackbody component derived from UV/optical fitting is plotted against those measured in the X-ray band. They are well consistent within errors, although there is an extrapolation over 3 orders of magnitude in wavelength. 

\section{Discussion}
\label{sec:discuss}

Compared with systems with a moderate accretion rate, a major difference for supercritical accretion is that an optically thick and nearly spherical wind will be launched, such that the manner of self-irradiation will be altered as a natural consequence.  In this work, we describe an irradiation model incorporating a RHD wind solution \citep{Meier1982_PaperII} coupled with a simple slim disk model \citep{Meier2012} to predict the UV/optical spectrum of the system, in an extreme that no hard X-rays from the central funnel shine the outer disk. This is an extension of the previous work of \citet{Zhou2019}, which focuses on the X-ray band.  

Here we note that, in theory, the slim disk model cannot be applied to neutron stars, because the advective power will be released on the hard surface and change the disk structure in turn.  As the wind launches at a large radius ($\dot{m} r_{\rm isco}$ and $\dot{m} \gg 1$), where the surface effect may be small, the model assumptions can still be valid if the compact object is a neutron star. However, if the neutron star is extremely magnetized and the disk truncation radius is larger than the wind launch radius, the model will not work any more.  

The model can adequately fit the UV/optical SED of NGC 247 X-1, with a median difference of 1.7\% and a maximum of 11\%, while fitting with a standard irradiation model results a median difference of 10\% and a maximum of 19\% \citep{Feng2016}. It is perhaps unfair to compare the goodness of the fit directly, because there are more free parameters in this model. However, the standard irradiation model requires a point-like irradiating source of about $10^{41}$~\ergs\ located at the center. This contradicts the X-ray observations, which suggest that the majority of the power arises from a blackbody component of a large radius. Plus, the bolometric luminosity derived from X-ray observations is also much lower than needed. Here, we provide a well self-consistent explanation for the UV/optical SED of NGC 247 X-1. The irradiating source, thermalized emission from the wind, is inferred to have a temperature of about 0.14~keV and a total luminosity of about $3.6 \times 10^{39}$~\ergs, well consistent with X-ray observations. We emphasize that this is an extrapolation over 3 orders of magnitude in wavelength. Such a coincidence is indeed a strong support of the model.  

The best-fit results suggest that there is a black hole with a mass of roughly 30~$M_\odot$ accreting at a rate of around a hundred times the critical value, with a scatter of 0.3~dex or so.  The inclination angle $i$ is constrained to be larger than 67\arcdeg, which is mainly determined by the ratio of flux at short and long wavelengths.  A high inclination is indeed expected from the observational fact that the source is supersoft in the X-ray band. Due to geometric beaming in the central funnel, an observer viewing the system at a high inclination sees very little amount of the hard X-ray emission, so does the outer disk. On the other hand, non-detection of eclipse, if is not due to insufficient observations, may suggest that the inclination cannot be extremely high. The critical angle for eclipse only depends on the mass ratio \citep{Eggleton1983}, and the above inclination constrains the secondary to compact object mass ratio to be less than 1.68, which is not extreme and allows both low or high mass companions if the compact object is a stellar mass black hole. For comparison, the Galactic black hole binary XTE~J1550-564 may have an inclination higher than 70 degrees \citep{Orosz2011}, but no eclipse has been observed. As the thermal emission leaves the system from the scattersphere, which has a radius of $8 \times 10^{11}$~cm or 12 $R_\odot$ given the best-fit results, partial eclipse will happen at a lower inclination angle. However, as the inferred scattersphere is much smaller than an evolved star, the finite size effect may be negligible.

Hard X-ray emission is usually the dominant component in standard ULXs. Although there is some degree of geometric beaming, the angular distribution of the hard X-rays could still be wide due to a large scattering optical depth and consequently a high-altitude scattersphere in the funnel \citep{Jiang2014}. In these cases, the self-irradiation becomes more complicated, as both the hard X-rays from the central funnel and the thermalized soft X-rays from the wind can illuminate the outer disk. This is the reason why a supersoft ULX is selected in this work for the purpose of a safe test of the model. To model irradiation in standard ULXs, one needs to add a point-like source at the center (to be precise, at the scattersphere of the funnel) to account for the heating resulted from hard X-rays. This would require high-quality UV and optical data to disentangle the degeneracy between the two irradiation sources.  Once this is done, the angular distribution or the beaming factor of the hard X-rays can be constrained. 
 
\acknowledgements We thank the anonymous referee for useful comments that help improve the paper. HF acknowledges funding support from the National Key R\&D Project (grants Nos.\ 2018YFA0404502 \& 2016YFA040080X), and the National Natural Science Foundation of China under the grant Nos.\ 11633003 \& 11821303. 



\end{document}